\documentclass[lettersize,journal]{IEEEtran}
\usepackage{amsmath,amsfonts}
\usepackage{algorithmic}
\usepackage{algorithm}
\usepackage{array}
\usepackage[caption=false,font=normalsize,labelfont=sf,textfont=sf]{subfig}
\usepackage{textcomp}
\usepackage{stfloats}
\usepackage{url}
\usepackage{verbatim}
\usepackage{graphicx}
\usepackage{cite}
\usepackage{mathptmx}
\usepackage{etoolbox}
\usepackage{lineno}
\usepackage{booktabs}
\usepackage{makecell}
\begin{document}

\title{Velocity-based sparse photon clustering for space debris ranging by single-photon Lidar}

\author{Xialin Liu, Jia Qiang, Genghua Huang, Liang Zhang, Zheng Zhao, Rong Shu
\thanks{The authors are with Key Laboratory of Space Active Opto-electronics Technology, Shanghai Institute of Technical Physics, Chinese Academy of Sciences, 500 Yutian Road, Shanghai 200083, China (liuxialin@mail.sitp.ac.cn, qiangjia@mail.sitp.ac.cn, genghuah@mail.sitp.ac.cn, zhliang@mail.sitp.ac.cn, 	zhaozheng@shanghaitech.edu.cn, shurong@mail.sitp.ac.cn)}
\thanks{X. Liu, J. Qiang, G. Huang, L. Zhang and R. Shu are with Shanghai Quantum Science Research Center, Shanghai 201315, China}
}



\maketitle

\begin{abstract}
Single-photon Lidar (SPL) offers unprecedented sensitivity and time resolution, which enables Satellite Laser Ranging (SLR) systems to identify space debris from distances spanning thousands of kilometers. However, existing SPL systems face limitations in distance-trajectory extraction due to the widespread and undifferentiated noise photons. In this paper, we propose a novel velocity-based sparse photon clustering algorithm, leveraging the velocity correlation of the target’s echo signal photons in the distance-time dimension, by computing and searching the velocity and acceleration of photon distance points between adjacent pulses over a period of time and subsequently clustering photons with the same velocity and acceleration. Our algorithm can extract object trajectories from sparse photon data, even in low signal-to-noise ratio (SNR) conditions. To verify our method, we establish a ground simulation experimental setup for a single-photon ranging Lidar system. The experimental results show that our algorithm can extract the quadratic track with over 99\% accuracy in only tens of milliseconds, with a signal photon counting rate of 5\% at -20 dB SNR. Our method provides an effective approach for detecting and sensing extremely weak signals at the sub-photon level in space.
\end{abstract}


\section{Introduction}
In recent years, the amount of space debris objects in orbit has increased exponentially, which threatens the safety of the operational spacecraft (Jason-1\cite{jason-1}, March 2002; Irid-ium33 collision with Cosmos 2251 in 2009) and poses a potential collision risk to future spacecraft launches, spacecraft testing, and other space activities. Space Debris Laser Ranging (DLR) is developed from Satellite laser ranging (SLR)\cite{greene2002laser}, and it is one of the most efficient measurement techniques to range the non-cooperative targets\cite{kucharski2014attitude}. Single-photon Lidar (SPL)\cite{5340610}, with unprecedented sensitivity and picosecond time resolution, can greatly improve the detection efficiency of ranging systems and reduce the requirement for laser pulse energy. However, the increased sensitivity to weak photon signals comes at the cost of inducing a large amount of background noise. A high-precision acquisition, tracking, and pointing (ATP) device in DLR system, along with accurate orbit prediction information\cite{li2016real,9393372}, can mitigate the impact of background noise and reduce the amount of data to a certain extent. Nevertheless, for sparse photon signals submerged in extensive and indistinguishable noise, fast and precise data processing methods are urgently needed for DLR\cite{Li19Ranging}.

Degnan et al. developed a signal extraction algorithm based on O-C residual comparison in SLR2000\cite{degnan2002optimization}. This algorithm divides the O-C residual plane into grids of equal size and compares the number of points in the grid to a set threshold to determine if the points are signal points. Automatic observation data processing algorithms based on Poisson filtering have also been implemented\cite{rodriguez2016assessing,garcia2020trajectory}. These methods assume that the noise points in the O-C residuals follow a Poisson distribution, and the signal photon echo residual falls on a straight line with an unknown slope over a short time interval. Hough Transform can be potentially a solution and has been widely used in line recognition\cite{mukhopadhyay2015survey}, and Liu et al. proposed an effective echo trajectory extraction algorithm based on random Hough transform\cite{tong2016effective}. This algorithm can effectively deal with the situation in which the signal is sparse and submerged by noise. However, projecting the echo photon signal into the high-dimensional parameter space introduces high spatial complexity and time complexity to the algorithm. 

In this study, we present an innovative velocity-based sparse photon clustering (VBSPC) algorithm for extracting target trajectories through single-photon ranging lidar. It utilizes the time-distance correlation of moving space debris targets, by calculating the velocities of photon distance points between consecutive pulses over a defined temporal span, clustering these photons based on their velocities, and fitting the precise target trajectories through the velocity-clustered photons. This method resolves the poor accuracy and sparse data invalidation found in direct measurement and statistical methods, while also avoiding the accuracy loss and substantial computation costs associated with rasterization in the image processing methods. To evaluate the feasibility and effectiveness of our algorithm, we built a ground simulation experimental system. In the system a 17 meters collimator is implemented to simulate transmitting a beam through thousands of kilometers, and the original echo photon data is acquired using SPADs and a high-resolution time-correlated single photon counting (TCSPC)\cite{wallace20013d} module. The processing results of the original photon data show that our method can successfully extract the motion trajectories for uniform speed, uniform acceleration, variable acceleration, and multiple targets within milliseconds while maintaining millimeter-level ranging accuracy. Our algorithm can achieve more than 99\% accuracy for a quadratic track with a signal photon detection efficiency of 5\% at -20 dB SNR. Our research provides an effective data processing approach for debris ranging using single-photon lidar systems.  
\section{Methods}

\begin{figure}[htbp]
\centering
\includegraphics[width=8.5cm]{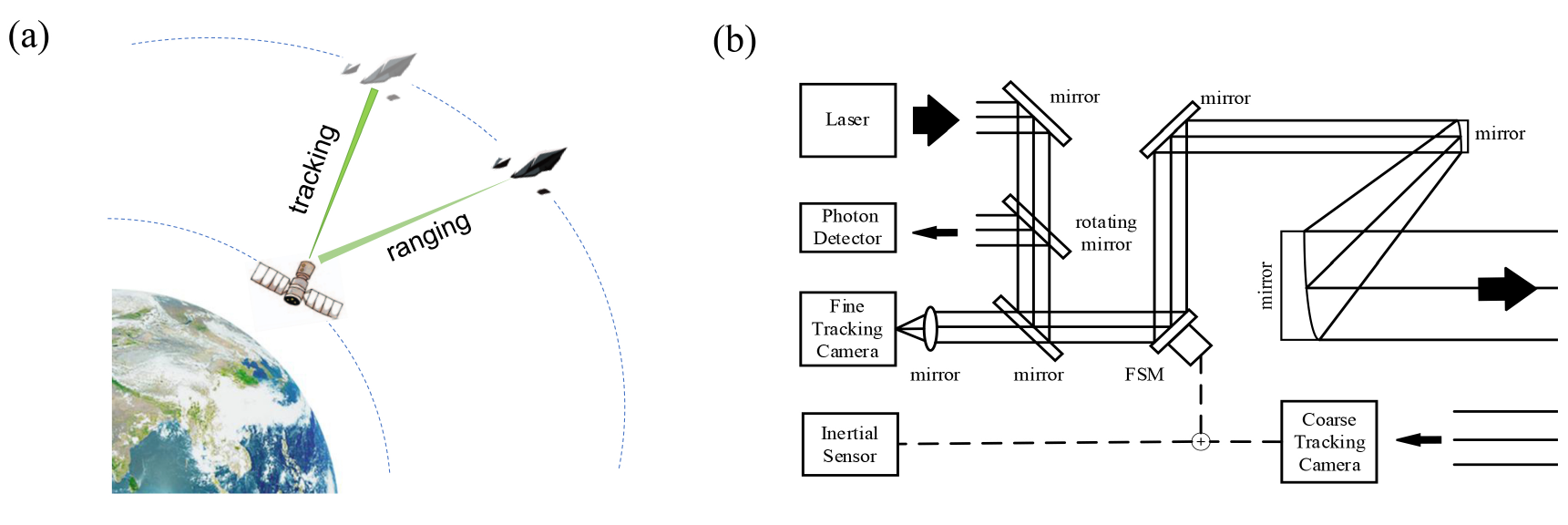}
\caption{(a) Conceptual diagram of laser ranging for non-cooperative debris. (b) Schematic diagram of laser ranging and tracking system.}%
\label{fig:schematic}
\end{figure}

In Fig.\ref{fig:schematic} (a) we depict the conceptual diagram of laser ranging for non-cooperative debris; Fig.\ref{fig:schematic} (b) displays the schematics of the ranging and tracking system. The system includes an ATP subsystem and a laser ranging subsystem that share the optical path. The ATP subsystem comprises a coarse tracking camera, a fine tracking camera, an inertial sensor, a piezoelectric deflection mirror (FSM) and other tracking actuators. The ranging subsystem mainly consists of a pulsed laser, a single photon detector, a time synchronization control and photon counting circuit modules.  After the tracking subsystem completes the acquisition and tracking of the non-cooperative debris and, the laser emits a pulse to illuminate the space target once the tracking accuracy meets the requirements of laser ranging; the diffused reflection echo signal is then collected by a telescope system and detected by an photon detector. The TCSPC module records the time difference between the laser pulse emission and the echo photon detection, which we use to calculate the original ranging value.

According to the photon Lidar equation, in the case where the debris at a long distance can be considered as a point target, the average returned photon numbers can be expressed by,
\begin{equation}
N_r = \left( \frac{\eta_q}{h\nu} \right) \frac{\eta_t\eta_r\tau^2cos(\theta_r)\gamma A_r A_t}{\pi \alpha^2 R^4}E_t
\label{eq:Nr}
\end{equation}
where $R$ is the distance between the Lidar and the target, $\alpha$ denotes the divergence angle, $\theta_t$ denotes the included angle of radar cross-section, $\eta_t$ and $\eta_r$ denote transmitted efficiency and receiving efficiency, $E_t$ is the emitted energy, $\gamma$ denotes the average reflectivity of the object, $\tau$ is the transmittance of the transmission medium, $A_t$ and $A_r$ represent the target and receiving area, $\eta_q$ represents the quantum efficiency of SPAD, $\nu$ is the frequency of light, and $h$ is the Planck constant.  

In low-light conditions,  the detection of photon signals, as well as the shot noise and background noise, follow independent Poisson random processes. The photodetection resulting from the superposition of these three processes remains a compound Poisson random process. The probability of the detector detecting k photon counts generated by the echo of a single pulse is:
\begin{equation}
P\{K=k\} = \frac{(N_r)^k e^{- N_r}}{k!}.
\label{eq:Pk}
\end{equation}

\begin{figure}[htbp]
\centering
\includegraphics[width=8.5cm]{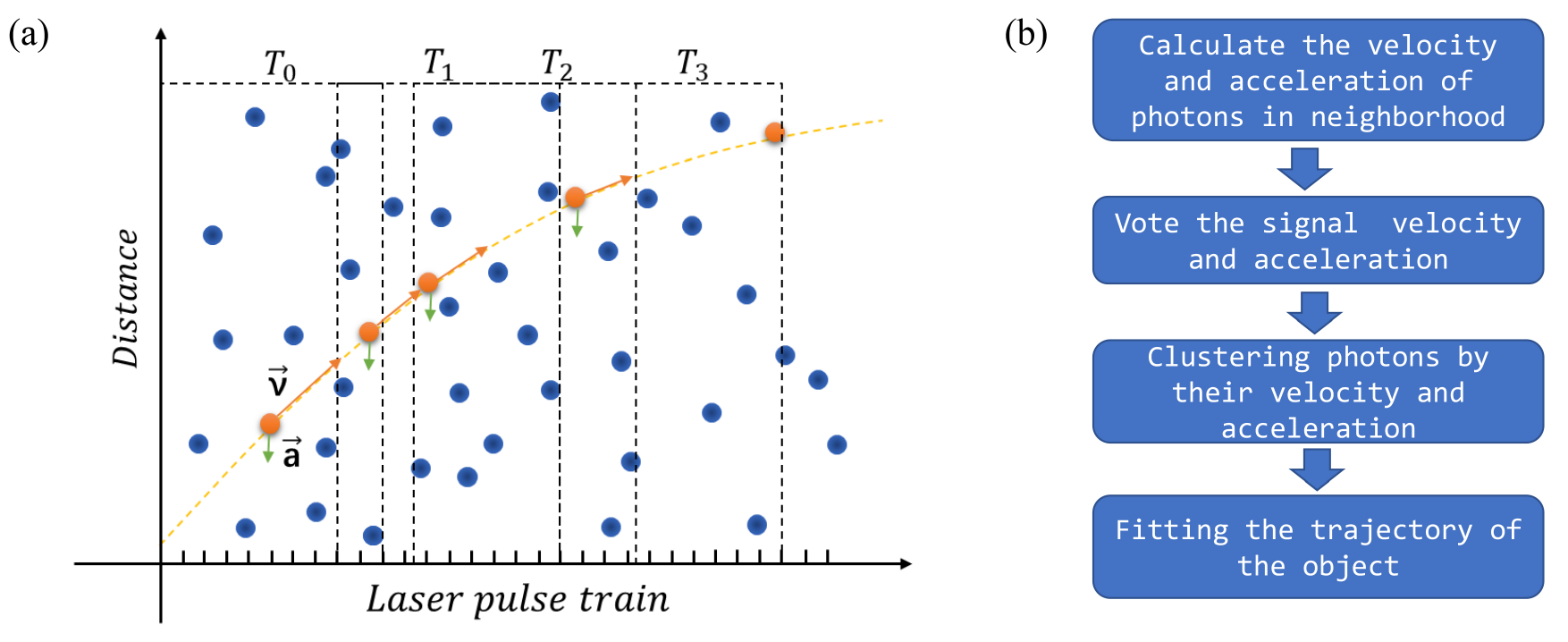}
\caption{(a) Conceptual diagram of the velocity-based sparse photon clustering algorithm. The yellow dots along the trajectory represent signal photons, and the blue dots represent noise photons. (b) Flow chart of the velocity-based sparse photon clustering algorithm.}%
\label{fig:algorithm}
\end{figure}

When $P\{K>1\} \ ll 0$, SPAD works in linear mode, and we can consider the probability of detecting a photon as,
\begin{equation}
P\{k=1\} \approx P\{K>0\} = 1-P\{k=0\}=1- e^{- N_r}.
\label{eq:P1}
\end{equation}
Satellites traverse space under the gravitational influence of various planets. Adhering to Kepler's Law, the orbital motion of satellites is predominantly elliptical with the Earth as the central focus [11], as illustrated in Fig. \ref{fig:schematic}. The distance between the target from the Lidar can be approximated as a second-order polynomial,
\begin{equation}
R = at^2+ bt +c, 
\label{eq:Rt}
\end{equation}
where $R$ is the ranging distance of the target from the Lidar, $t$ is the ranging time, $a$, $b$, and $c$ are parameters of the polynomial which are decided by the running orbits. The photon detection distribution of space debris ranging is illustrated in Fig.\ref{fig:algorithm} (a), where the yellow dots along the trajectory represent signal photons, and the blue dots represent noise photons.

Fig.\ref{fig:algorithm} (a) also shows the conceptual diagram of the velocity-based sparse photon clustering with high background noise. The distance of the $j$th photon collected at time $t_i$ is denoted as $(t_i, R_{ij})$. The average time interval between the detection of two signal photons is defined as the time neighborhood $\epsilon$, a value correlated with the signal photon detection probability as shown in Eq.\ref{eq:P1}. We identify signal photons by computing and tallying the velocities and accelerations of photons within a segmented search duration $T$. Fig.\ref{fig:algorithm} (b) shows the conceptual diagram and flow chart of the VBSPC algorithm. The velocity and acceleration of the $j$th photon collected at time $t_i$ relative to the $j'$th photon collected at time $t'_i$ are calculated by $\vec{v}_{ij}=\frac{R_{ij}-R_{i'j'}}{t_i-t_{i'}}$ and $\vec{a}_{ij} = \frac{v_{ij}-v_{i'j'}}{t_i-t_{i'}}$. The detailed process of the algorithm is outlined in Algorithm 1. Here, $T_d$ represents the detection duration, hence the width of segmented search duration $T$ is less than $T_d$. The absolute difference of velocities and accelerations within the neighborhood, $\|\vec{v}_{ij}-\vec{v}_{i'j'} \|$ and $\|\vec{a}_{ij}-\vec{a}_{i'j'} \|$ respectively, is determined. By setting thresholds $\delta_v$ and $\delta_a$ for the differences in velocity and acceleration between photons, we distinguish between signal and noise photons. These thresholds are influenced by the actual motion velocity of the target and the noise level, which can be obtained through specific data statistics. After clustering signal photons based on speed and acceleration, we perform a polynomial fitting on the ranging data of photons in the set $\left\{C_k \right\} $, thereby obtaining the trajectory of the target.

\begin{algorithm}[H]
\caption{ Velocity-based sparse photon clustering algorithm.}
\label{alg:VBSPC}
\begin{algorithmic}[1] 
\REQUIRE ~~\\ 
DataBase, $D = \left\{(t_i,R_{ij}) \right\}$;\\
The segmented search duration, $T = \epsilon$;\\
\ENSURE ~~\\ 
Signal photon cluster, $\left\{C_k \right\} $
Fitted trajectory, $\left\{y_k \right\}$;
\STATE Calculating the velocity $\left\{\vec{v}_{ij} \right\} $ and the acceleration $\left\{\vec{a}_{ij} \right\} $ of photons in the search duration $T$;
\label{ code:Calculating }
\STATE Searching the velocity and the acceleration of photons in the segmented search duration $T$. If $\|\vec{v}_{ij}-\vec{v}_{i'j'} \|<\delta_v$ or $\|\vec{a}_{ij}-\vec{a}_{i'j'} \|<\delta_a$, add $(t_i,R_{ij})$ to set $\Omega$;
\label{code:Searching}
\STATE If $\Omega = \emptyset$, increase the search depth $T = 2T$, and return to step \ref{code:Searching}. If $T = T_d$ and $\Omega = \emptyset$, terminate the algorithm;
\label{code:add}
\STATE Voting on the speed and acceleration of the core points in $\Omega$, and clustering the points that exceed the noise threshold and have the same speed or acceleration into $\left\{C_k\right\}$;
\label{code:classify}
\STATE Fitting the motion trajectories $y_k$ with points in $C_k$;
\label{code:fitting}
\RETURN $\left\{C_k \right\}, \left\{y_k \right\}$; 
\end{algorithmic}
\end{algorithm}

\begin{figure}[htbp]
\centering
\includegraphics[width=8cm]{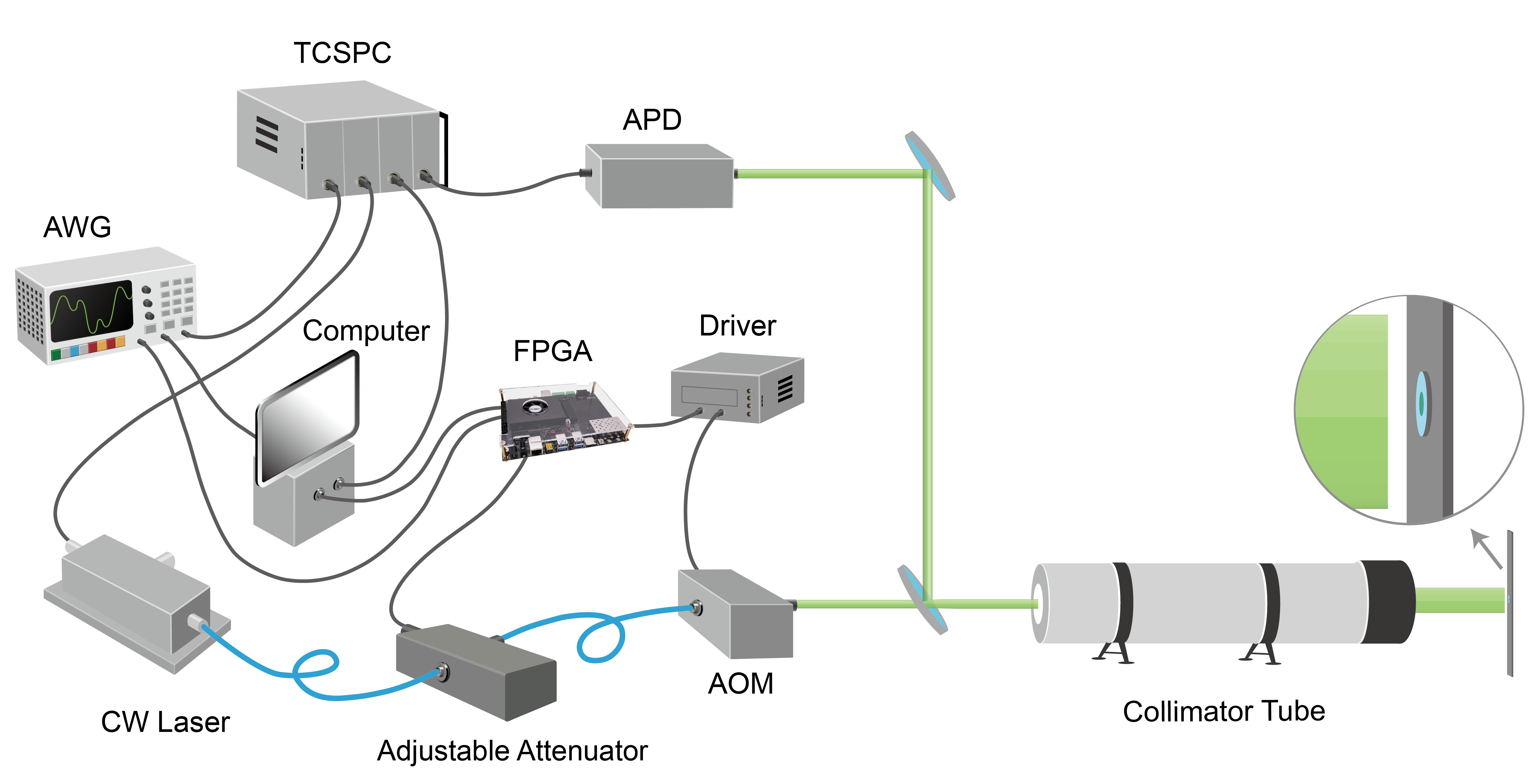}
\caption{Experimental setup diagram of the ground simulator of SPL.}%
\label{fig:Gsimulator}
\end{figure}

To verify the feasibility of our algorithm, we built an SPL ground detection simulator as shown in Fig.\ref{fig:Gsimulator}. A 17-meter-long collimator is used to simulate the long-distance transmission of the light field. An acousto-optic modulator (AOM) is implemented to perform high-precision pulse cutting and modulation of a continuous-wave (CW) laser. An adjustable attenuator is used to control the intensity of light. The Arbitrary Waveform Generator (AWG) is used to generate a sequence signal representing a certain frequency of laser pulses, which will be synchronized with the photon detection signal through TCSPC. An FPGA programmable control module is incorporated to control the width, delay, and intensity of the echo laser pulses, which simulate the echo photon data under different laser pulse widths, target distance, and reflectivity in a long-range SPL ranging system.

\section{Results}

\begin{figure}[htbp]
\centering
\includegraphics[width=8.5cm]{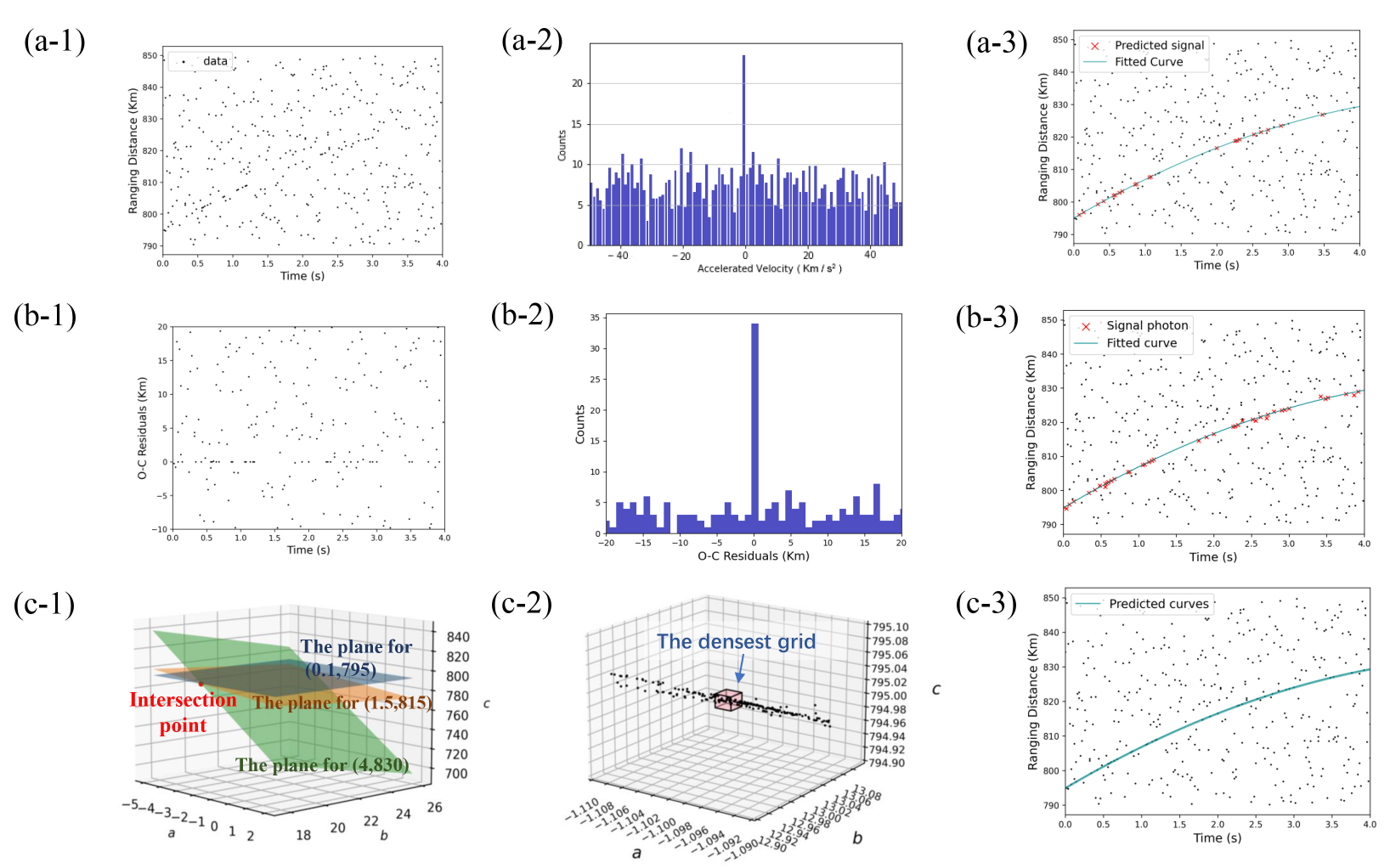}
\caption{The comparison of three methods, VBSPC (a-1,2,3), O-C residuals (b-1,2,3), and Hough transform (c-1,2,3), for extracting quadratic motion trajectory from photon ranging data.}%
\label{fig:process}
\end{figure}

Fig. \ref{fig:process} illustrates the process of extracting sparse photon signals for quadratic trajectory targets using three methods, VBSPC (a-1,2,3), O-C residuals (b-1,2,3), and Hough transform (c-1,2,3). The photon data was obtained from the SPL ground detection setup as shown in Fig. \ref{fig:Gsimulator}.  The laser repetition rate is 50 Hz, the signal photon counting rate is 10\%, and the signal-to-noise ratio (SNR) is -13 dB. 
The photon counting rate, denoted as $P_s$,  is defined as the average number of signal photons from individual laser pulse echoes, equivalent to the photon detection probability in Eq.\ref{eq:P1}. Fig. \ref{fig:process} (a-1) shows the original photon-ranging data in which signal photons are submerged by noise photons. By calculating and counting the photon's velocity and acceleration in the segmented search duration, we obtained the acceleration value of signal photons, which is significantly higher than the noise photon acceleration, as shown in Fig. \ref{fig:process} (a-2). We marked those photons with the right acceleration as the signal photons, and fitting these signal photon data points gives us the target trajectory, as shown in Fig. \ref{fig:process} (a-3). Fig. \ref{fig:process} (b-1) shows the O-C residual plot obtained by subtracting the raw photon ranging data from the expected trajectory data. Here, we use the accurate target trajectory as the predicted trajectory. Histogram statistics are applied to the residuals as shown in Fig. \ref{fig:process} (b-2), and photons from the peak histogram bin are selected as signal photons. Fitting these signal photons allows us to obtain the target trajectory. While this method allows for the rapid identification of signal photons, it fails to distinguish between signals and noise within the same histogram bin. As shown in Fig. \ref{fig:process} (b-3), this approach tends to misclassify some noise photons as signals. Furthermore, the substantial deviation in the predicted trajectory data under real-world conditions contributes to an even larger margin of error in the practical application of this method. The basic idea of the Hough Transform is to represent curves in parameter space. For a quadratic curve, one photon data point in the original plot Fig. \ref{fig:process} (a-1) corresponds to a plane in the parameter space. The intersection of three planes represents a set of parameters (a, b, c) that can be obtained from three sets of photon data, as illustrated in Fig. \ref{fig:process} (c-1). Permuting and combining the original data to calculate all the intersection points in the parameter space, and then discretizing the parameter space and identifying the cell with the highest count of intersections, as shown in Fig. \ref{fig:process} (c-2). The parameter values corresponding to this bin are the sought-after parameters for the target trajectory as shown in Fig. \ref{fig:process} (c-3). The size of the discretization grid determines the precision of parameter estimation. If the grid is set small enough, the Hough Transform can achieve high accuracy and precision, but at the cost of significantly increased computational time. 

To analyze the performances of these three methods, we use Recall to emphasizes the model's ability to identify all relevant instances, Accuracy to provide a general measure of correctness across all classes, and Precision to highlight the accuracy of positive predictions. We categorize photon data into true positives (TP), false positives (FP), true negatives(TN), and false negatives (FN) based on their actual signal and noise categories, combined with the algorithm's predicted categories. Therefore, Recall, Accuracy, and Precision can be defined as:
\begin{equation}
Recall = TP / (TP + FN)
\label{eq:recall}
\end{equation}
\begin{equation}
Accuracy = (TP + TN) / (TP + FP + TN + FN)
\label{eq:accuracy}
\end{equation}
\begin{equation}
Precision = TP / (TP + FP)
\label{eq:precision}
\end{equation}

\begin{table}[htbp]
    \footnotesize
    \centering
    \caption{The comparison of performances between the existing algorithms and ours.}
    \label{tab:alg-compa}
    \begin{tabular}{l|cccccc}
    \toprule
        \textbf{Algorithms} &\textbf{\makecell{Recall}} & \textbf{Accuracy} & \textbf{Precision} & \textbf{Time (ms)} \\ \midrule
        O-C residuals\cite{degnan2002optimization,rodriguez2016assessing} & 0.94 & 0.92 & 0.71 & 1 \\
        Hough transform\cite{tong2016effective} & 0.95 & 0.85 & 0.89 & 980 \\
        Our algorithm & 0.85 & 0.97 & 0.96 & 5 \\
    \bottomrule
    \end{tabular}
\end{table}
We conducted ten experiments under the same parameter conditions and calculated the average values for each performance metric of each algorithm. The comparisons are shown in Table \ref{tab:alg-compa}. We can see that the O-C residuals algorithm can handle the spatially sparse photon ranging data with relatively low computing complexity, but it has limited precision and is unable to deal with situations where there is a lack of trajectory prediction or when the predicted trajectory has a significant deviation. Hough transform algorithm offers a satisfactory performance under sparse signal situations. However, its space complexity scales unfavorably with precision, and it exponentially grows with an increase in the number of parameters. Our algorithm could potentially solve the above issues, offering exceptional processing speed and high precision for sparse photon-ranging data. 

\begin{figure*}[htbp]
\centering
\includegraphics[width=13.5cm]{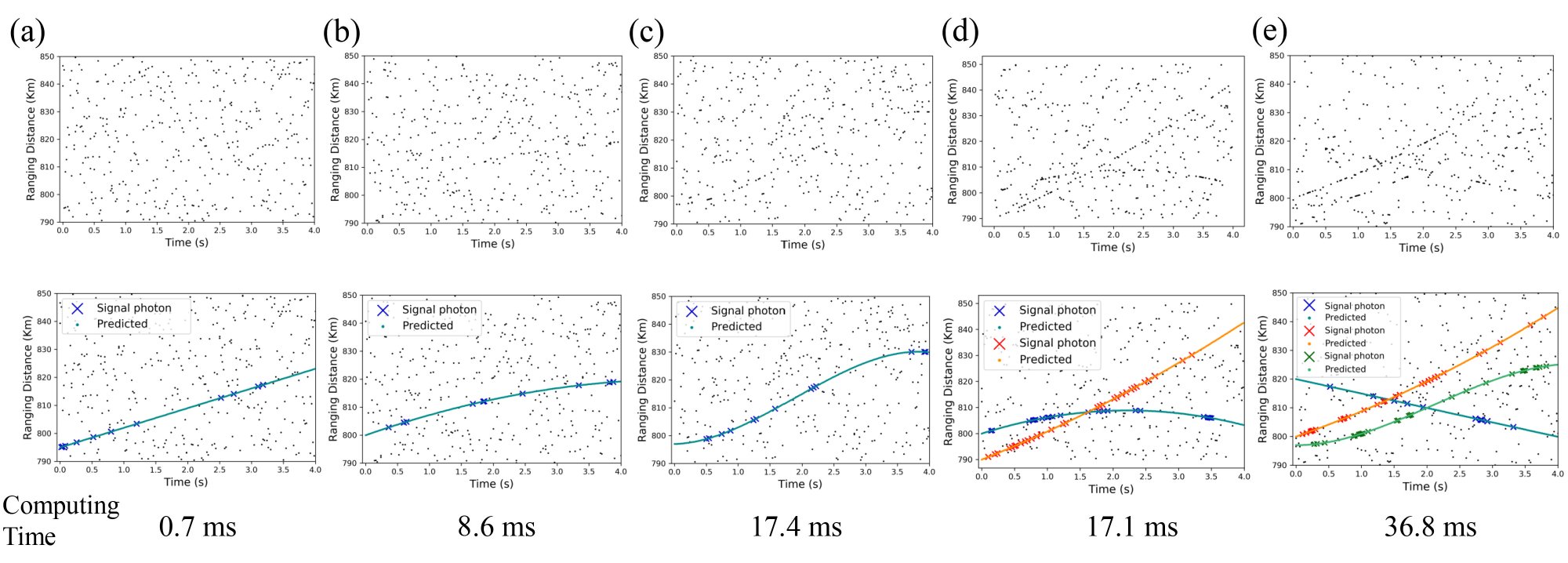}
\caption{The original photon data and the extraction results through the VBSPC algorithm for targets with different trajectories.}%
\label{fig:results1}
\end{figure*}

We simulated and detected photons of moving targets along different trajectories by the SPL simulator, and the original photon data and the extraction results are shown in Fig. \ref{fig:results1}. Fig. \ref{fig:results1} (a), (b) and (c) show the target detection and trajectory extraction results for straight, quadratic, and cubic curve trajectories, respectively. The required calculation time increases with the order of the curve as the amount of calculation required to solve the relevant values of velocity and acceleration increases. Fig. \ref{fig:results1} (d) and (e) show the photon data and relevant results for situations where two and three targets of different trajectories are measured simultaneously.  We can see from these results that the VBSPC algorithm has good recognition and trajectory extraction capabilities for signal photons from different trajectory targets and multiple targets under high background noise.

\begin{figure*}[htbp]
\centering
\includegraphics[width=14cm]{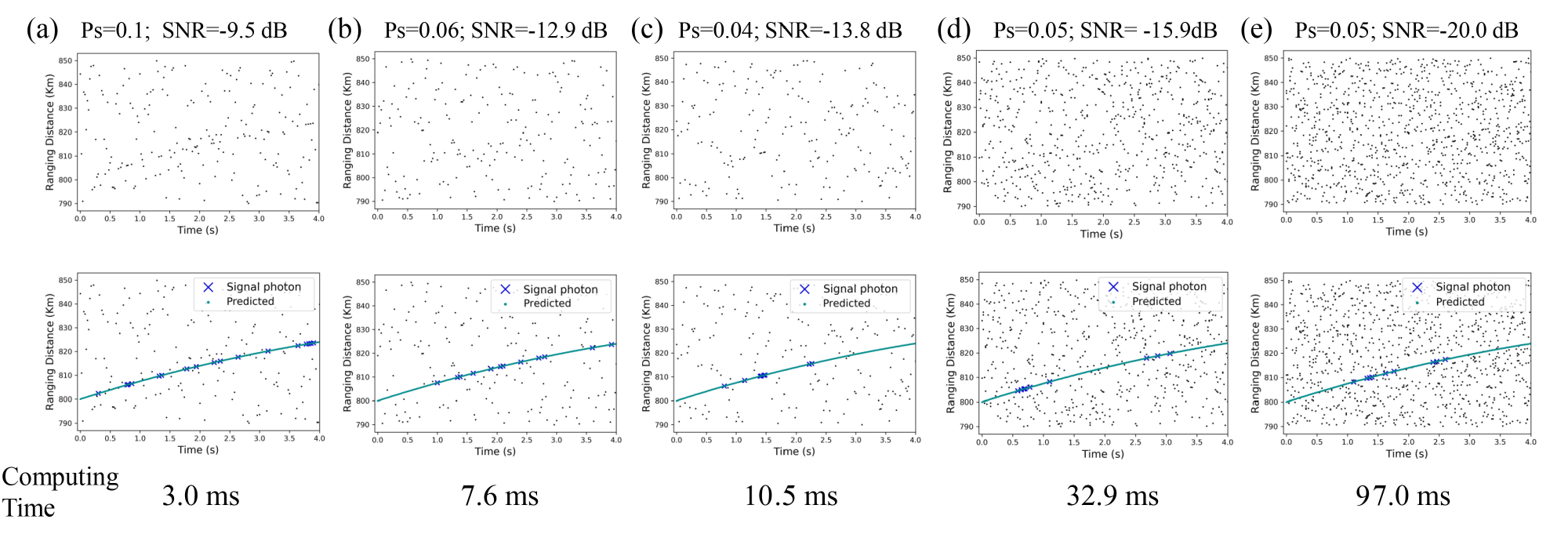}
\caption{The original photon data and the extraction results through VBSPC algorithm for a target with a quadratic curve trajectory under different photon counting rate ($P_s$) and SNR.}%
\label{fig:results2}
\end{figure*}
The original photon data and the extraction results for a target with a quadratic curve trajectory under different signal and noise photon counting rates are shown in Fig. \ref{fig:results2}. The signal photon counting rate, $P_s$, decreases from Fig. 7 (a) to (c), and the counting rate of noise photons increases from Fig. \ref{fig:results2} (c) to (e). The decrease in signal photon counting rate requires an increase in the segmented search duration $T$, resulting in longer computing time. On the other hand, a low SNR will also contribute to the increase in computation time since we will be searching among a larger number of photons in the segmented search duration. We can see from Fig. \ref{fig:results2} (e) that  our algorithm can still accurately find the submerged signal photons with the signal-to-noise ratio is as low as -20 dB.

\begin{figure}[H]
\centering
\includegraphics[width=8.5cm]{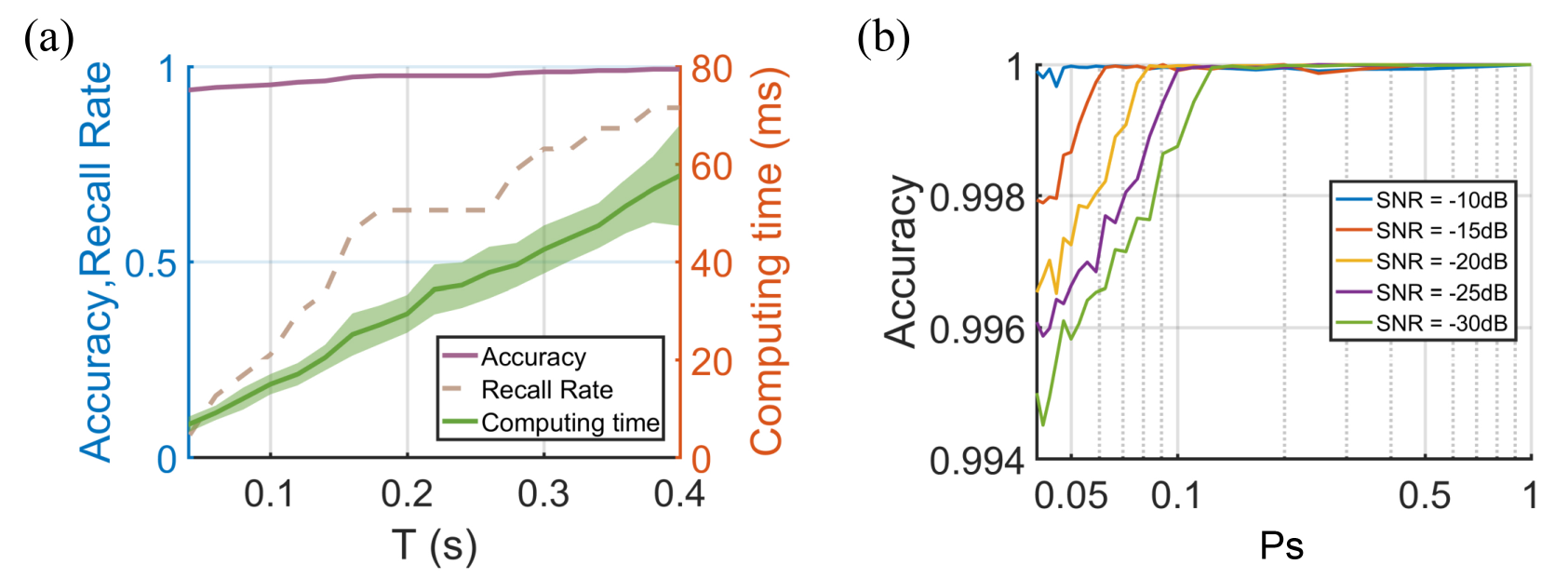}
\caption{(a) The accuracy and recall rate of VBSPC algorithm with different segmented search durations ($T$). (b) The accuracy versus signal photon counting rate ($P_s$) under different background noise rates.}%
\label{fig:accuracyVsT}
\end{figure}

In Fig. \ref{fig:accuracyVsT} (a),  we calculated the accuracy rate, recall rate, and computation time for a quadratic-curve target with the segmented search duration $T$ ranging from 0.02 s to 0.4 s at 50 Hz pulse repetition rate, 10\% signal photon counting rate, and -20 dB SNR. We can observe that as the segmented search duration $T$ increases, both the accuracy rate and recall rate improve. However, at the same time, the required computation time also increases. When the segmented search duration exceeds 0.3 s, we obtain high accuracy rates of over 99\% and recall rates of over 80\%. Each point in the plot corresponds to 100 simulations, with the lines indicating the mean value, while the shaded regions correspond to $\pm 1$ standard deviation. Each point in the plot corresponds to 100 simulations, with the lines indicating the mean value, while the shaded regions correspond to $\pm 1$ standard deviation. The changes in accuracy concerning signal photon counting rate at different SNRs are plotted in Fig. \ref{fig:accuracyVsT} (b). We can see that even with an SNR of -30 dB, that is, the noise photon count being 1000 times the signal photon count, our algorithm can still achieve an accuracy rate of over 99\%. We could not attempt experiments with lower SNR because excessive noise photon counts can cause overflow in the hardware readout circuit of the SPAD. From the principles and experimental results, it can be seen that noise has little impact on the accuracy of our algorithm, but it will increase the computing time required to search for signal photons.

\section{Conclusion}
In conclusion, we proposed a novel algorithm called VBSPC, capable of extracting signal photons reflecting from the targets moving along arbitrary polynomial curve trajectories in high background noise. A ground simulation experimental setup is established to verify our method. Our algorithm can achieve an accuracy of more than 99\% for a quadratic track with a signal photon counting rate of 5\% at -20 dB SNR. The computation time is about tens of milliseconds order, which is much faster than the Hough Transform algorithm\cite{tong2016effective}. With a relatively simple time and spatial complexity, our algorithm can be easily implemented in hardware processing systems. Our research provides an effective data processing approach for satellite or space debris ranging by single-photon Lidar.

\section*{Acknowledgments}
The authors acknowledge support through the Shanghai Pilot Program for Basic Research – Chinese Academy of Science, Shanghai Branch (JCYJ-SHFY-2021-04). X. L. acknowledges the financial support from the National Natural Science Foundation of China (62305360), Shanghai Science and Technology Development Funds (22YF1456000), and the Innovation Project of Shanghai Institute of Technical Physics (CX-366).



\section{References}

\bibliographystyle{IEEEtran}
\bibliography{IEEEabrv,myReference}


 





\end{document}